\documentclass[twocolumn]{aastex63}

\pdfoutput=1

\received{October 17, 2020}
\revised{December 15, 2020}
\accepted{December 19, 2020}
\submitjournal{ApJL}

\shorttitle{Flat and Fast Prograde Stellar Stream}
\shortauthors{Re Fiorentin et al.}

\begin{document}

\title{Icarus: a Flat and Fast Prograde Stellar Stream in the Milky Way disk}

\correspondingauthor{Paola Re Fiorentin}
\email{paola.refiorentin@inaf.it}

\author[0000-0002-4995-0475]{Paola Re Fiorentin}
\affil{INAF - Osservatorio Astrofisico di Torino, Strada Osservatorio 20, I-10025 Pino Torinese (TO), Italy}

\author[0000-0003-1732-2412]{Alessandro Spagna}
\affil{INAF - Osservatorio Astrofisico di Torino, Strada Osservatorio 20, I-10025 Pino Torinese (TO), Italy}

\author[0000-0003-0429-7748]{Mario G. Lattanzi}
\affil{INAF - Osservatorio Astrofisico di Torino, Strada Osservatorio 20, I-10025 Pino Torinese (TO), Italy}
\affil{Department of Physics, University of Torino, Via Giuria 1, I-10125 Torino, Italy}

\author[0000-0001-6291-6813]{Michele Cignoni}
\affil{Department of Physics, University of Pisa, Largo Pontecorvo 3, I-56127 Pisa, Italy}
\affil{INFN, Largo B. Pontecorvo 3, I-56127 Pisa, Italy}
\affil{INAF - Osservatorio di Astrofisica e Scienza dello Spazio, Via Gobetti 93/3, I-40129 Bologna, Italy}

\begin{abstract}
We explore the local volume of the Milky Way via chemical and kinematical measurements 
from high quality astrometric and spectroscopic data recently released by the Gaia, APOGEE and GALAH programs. 

We chemically select $1137$ stars up to $2.5$~kpc of the Sun and $\rm{[Fe/H]} \le -1.0$~dex, and find evidence of statistically significant substructures. 
Clustering analysis in velocity space classifies $163$ objects into eight kinematical groups, 
whose origin is further investigated with high resolution N-body numerical simulations of single merging events.

The two retrograde groups appear associated with Gaia-Sausage-Enceladus, while the slightly prograde group could be connected to GSE or possibly Wukong. 

We find evidence of a new 44-member-strong prograde stream we name Icarus; 
to our knowledge, Icarus is the fast-rotating stream closest to the Galactic disk to date 
($\langle Z_{\rm max} \rangle \lesssim 0.5$~kpc, $\langle V+V_{\rm{LSR}}\rangle \simeq 231~\rm{km~s^{-1}}$).
Its peculiar chemical ($\langle \rm{[Fe/H]}\rangle \simeq -1.45$, $\langle \rm{[Mg/Fe]}\rangle \simeq -0.02$) 
and dynamical (mean eccentricity $\simeq 0.11$) properties 
are consistent with the accretion of debris from a dwarf galaxy progenitor with a stellar mass of 
$\sim 10^9  M_\sun$ on an initial prograde low-inclination orbit, $\sim 10^\circ$.

The remaining prograde groups are either streams previously released by the same progenitor of Icarus (or Nyx),  
or remnants from different satellites accreted on initial orbits at higher inclination.

\end{abstract}
 


\section{Introduction}\label{sec:1}

According to current formation models,
galaxies like the MilkyWay (MW) grow by mergers of smaller satellites over their lifetime. Simulations based on this 
cosmological paradigm show 
that tidal forces can distort or even disrupt low-mass systems orbiting 
a MW analogue. This process rips out stars from the progenitors
leaving them as fossil debris with inhomogeneous distributions 
in the spheroidal  (halo-like) component of the host 
\citep[][]{Johnston1998, Bullock2005, Cooper2010, Fattahi2020}. 

Considerable structure is still present in the Galactic halo 
that does retain memory of its accretion history in the form of streams of stars \citep[e.g.][]{Ibata1994, Malhan2018, Naidu2020}. 
Besides, in the vicinity of the Sun (within, say, $3~\rm{kpc}$), where strong phase-mixing takes place, merger debris 
can still be identified as kinematical coherent streams 
despite of being spatially undetactable \citep[][]{Helmi1999, Smith2009, Klement2010, ReFiorentin2015}. 

Recent studies confirmed that a massive dwarf galaxy, 
named Gaia-Sausage-Enceladus (GSE), merged with the MW $\sim 10~\rm{Gyr}$ ago \citep[][]{Belokurov2018, Helmi2018, DiMatteo2019, Gallart2019}. 

More kinematical and chemical substructures with retrograde motions have been found among MW halo stars, i.e., 
Sequoia by \citet[][]{Myeong2019},  
Thamnos by \citet[][]{Koppelman2019}, and DTGs of \citet[][]{Yuan2020}.
Conversely, the prograde part of the halo has been little explored. 
Likely, the reason is the traditional kinematical selection criterions that reject as halo stars objects with 
$\vert\vert{\bf v}-{\bf v}_{\rm LSR}\vert\vert< v_{\rm lim}$, where $v_{\rm lim}=180\div 230~\rm{km~s}^{-1}$ 
\citep[][]{Nissen, Koppelman2019}.

The detection of accreted stars in the Galactic disk is more challenging, as these stars are 
difficult to distinguish being dominated by the {\it in situ} disk stars,  
even when chemical and dynamical informations are added \citep[][]{Ruchti}. 
Nevertheless, three new prograde streams were recently discovered: 
Nyx \citep[][]{Necib2020}, Aleph and Wukong \citep[][]{Naidu2020}.

Large unbiased (non-kinematically selected) samples of stars with accurate 6D phase-space information and 
chemical properties for classification and 
characterization 
can be obtained from high-precision data already (or soon to be) available. 

The Gaia second Data Release \citep[DR2;][]{GaiaCollaboration2018} 
provides unprecedented accurate
measurements of parallax and proper motion 
for more than $1.3~\rm{billion}$ stars across the whole sky. 
The Apache Point Observatory Galactic Evolution Experiment 
\citep[APOGEE~DR16;][]{Majewski2017, Ahumada2019} 
and the Galactic Archaeology with HERMES 
\citep[GALAH~DR2;][]{DeSilva2015,Buder2018}
have contributed 
high-resolution ($R \sim 22\,500$ near-infrared and $R \sim 28\,000$ optical, respectively) spectra   
yelding precise radial velocities, stellar parameters
and abundances for more than $20$ chemical elements. 

Here, we exploit the excellent synergy between the aforementioned surveys, and 
take advantage of these high-quality data 
to study chemo-kinematical signatures in the local halo, 
with particular attention to finding and characterizing accreted material towards the disk.


\section{Data and sample selection}\label{sec:2}

\begin{figure}
	\centering
	\includegraphics[width=\linewidth]{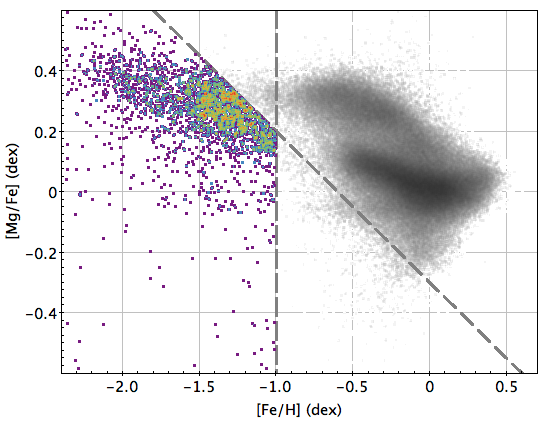}\\
	\includegraphics[width=\linewidth]{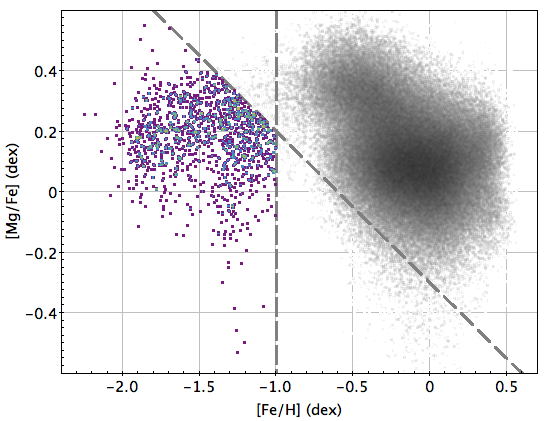} 
	\caption{
	{\bf Top:} Chemical distribution, $\rm{[Mg/Fe]}$--$\rm{[Fe/H]}$, for the $247\,013$ Gaia-APOGEE stars. 
	The dashed lines represent the adopted selection of $2517$ halo stars (color), separated from the thick/thin disk stars (gray). 
	{\bf Bottom:} Same distribution as top panel for the $1264$ halo stars selected among the $190\,559$ Gaia-GALAH stars.
	}
	\label{fig:fig1}
\end{figure}

This study starts with assembling a chemo-kinematical catalog 
by cross-matching Gaia~DR2,  APOGEE~DR16 and GALAH~DR2. 

It contains Gaia positions, parallaxes and proper motions \citep[][]{Lindegren2018a}, 
plus radial velocities and chemical abundances derived with the 
APOGEE and GALAH stellar spectra parameters pipelines \citep[e. g.,][]{Holtzman2015,Garcia2016,Kos2017,Buder2018}. 
In case of multiple spectroscopic observations, we adopt the one with highest nominal SNR. 

Firstly, we select objects having RUWE\footnote{RUWE stands for Renormalised Unit Weight Error.} 
$\le 1.4$, as extracted from Gaia.  
This, to discard sources with problematic astrometric solutions, 
astrometric binaries and other anomalous cases \citep[][]{Lindegren2018b}. 
Next, we retain only stars with three Gaia photometric bands and 
relative parallax error $\varpi/\sigma_\varpi >5$ (i.e., inverse-parallax distances better than $20\%$) 
for a total of $578\,976$ objects.

For the selection of sufficiently good APOGEE spectra, 
we reject stars with $\rm{SNR}<50$ and $\chi^2 < 25$ \citep[][]{Queiroz2020}. 
As for the GALAH data, 
we remove stars with flags warning of poor stellar parameters, and those with $\rm{SNR}<20$, as per \citet[][]{Hayden2020}. 

Therefore, we are left with a ``science'' sample of $437\,572$ stars down to $G=18~\rm{mag}$. 
Median uncertainties are: $0.03~\rm{mas}$ in parallax, 
$\sim 50~\rm{\mu as~yr^{-1}}$ in proper motion,  
and $\sim 40~\rm{m~s}^{-1}$ in line-of-sight velocity for the 
$247\,013$ Gaia-APOGEE and $190\,559$ Gaia-GALAH sources.

Galactic coordinates\footnote{We employ right-handed frames of reference with the axes positive toward the Galactic center, in the direction of Galactic rotation, and toward the North Galactic Pole, respectively.}  
and velocity components are derived by assuming that the Sun is $8.2~\rm{kpc}$ away from the MW centre,  
LSR rotates at $V_{\rm{LSR}}=232~\rm{km~s}^{-1}$ around the Galactic centre \citep[][]{McMillan2017}, 
and the LSR peculiar velocity of the Sun is $(U,V,W)_{\sun}=(11.1,12.24,7.25)~\rm{km~s}^{-1}$ \citep[][]{Schoenrich2010}. 
Median uncertainties of the resulting Galactocentric velocities are below $0.5~\rm{km~s}^{-1}$ for each component.

We also compute the orbital parameters of each entry (e.g., eccentricity and $Z_{\rm max}$) 
by adopting the Galactic potential model \verb+MWPotential2014+ from \citet[][]{Bovy2015}.

Figure~\ref{fig:fig1} shows the chemical plane, $\rm{[Mg/Fe]}$--$\rm{[Fe/H]}$, for the full chemo-kinematical catalog\footnote{
The contribution of APOGEE and GALAH is shown separately. 
Clearly, individual elemental abundances, derived with the multistep approach ``SME+The Cannon" for the GALAH spectra, 
are underestimated for low metallicity objects. 
Nevertheless, this survey remains appropriate to select halo tracers. 
 }.
Clearly, the sample is dominated by thin and thick disk stars.
We chemically identify halo stars 
by taking objects with $\rm{[Fe/H]} < -1.0$ 
and with $\rm{[Mg/Fe]}$ according to the relation: 
\begin{equation}
\rm{[Mg/Fe]} < -0.2 -0.5\cdot (\rm{[Fe/H]+0.2)},
\label{eq}
\end{equation}
we derived from \citet[][]{Mackereth2019}. 

We emphasize that this selection allows us to look for the metal poor component of accreted streams with disk-like kinematics right down the Galactic plane. 
Alternatively, prograde streams can be also detected using stellar samples selected {\it above} the plane without any metallicity cuts \citep[e.g., $|Z|>2~\rm{kpc}$ as in][]{Naidu2020}; 
however, such methodology cannot clearly identify accreted debris with thin disk-like kinematics.

In the kinematical analysis below we further remove known members of globular clusters, dSph 
\citep[e.g.][]{GaiaCollaboration2018-GC}, visual binaries and common proper motion pairs.
This last selection leaves  $3781$ halo stars up to $10~\rm{kpc}$. 

\begin{figure*}
	\centering
	\gridline{\fig{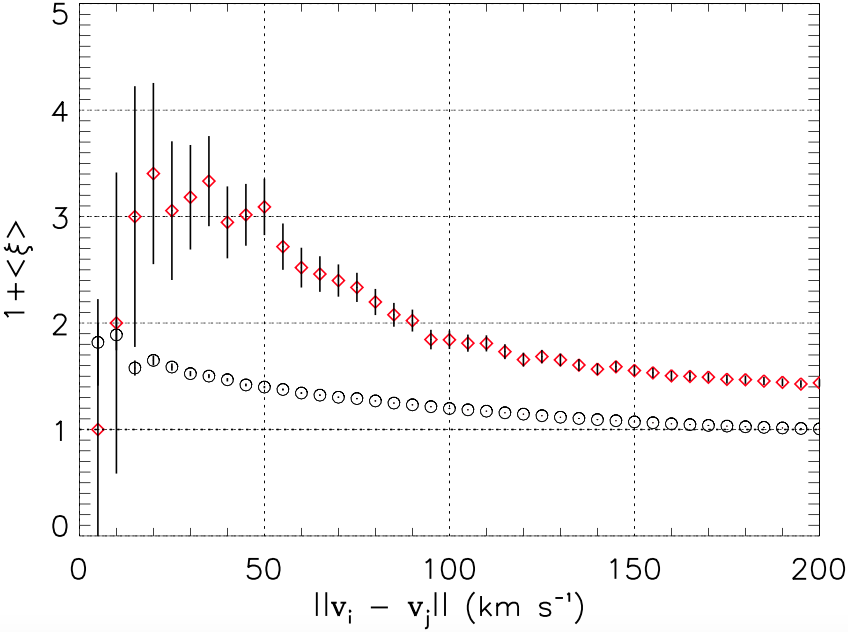}{0.36\textwidth}{}
                      \fig{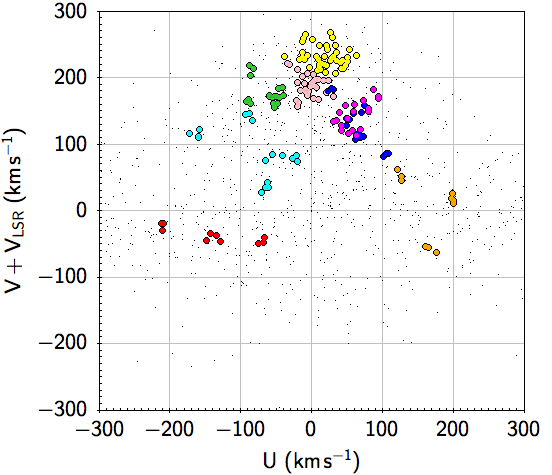}{0.32\textwidth}{}
                      \fig{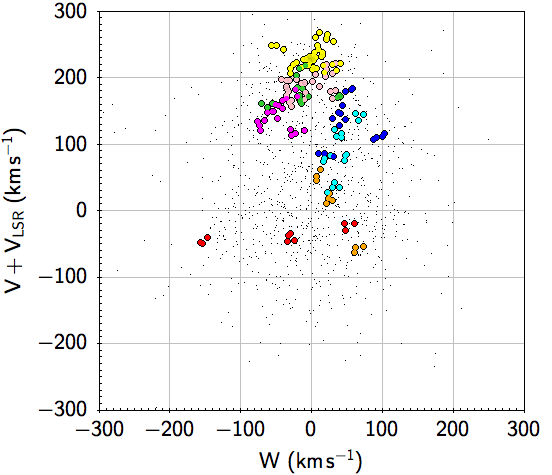}{0.32\textwidth}{}
                      }
        	\gridline{\fig{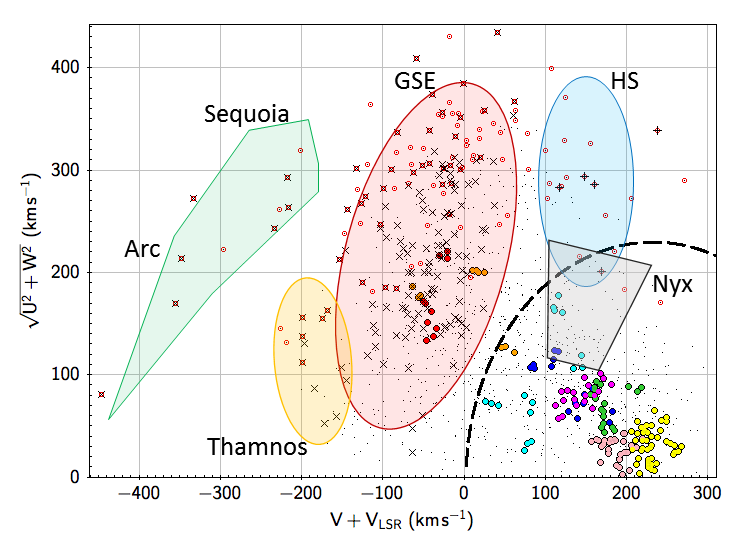}{0.7\textwidth}{}
                      }              
	\caption{
	{\bf Top left:} Cumulative velocity correlation function for the 
	$1137$ chemically selected nearby ($d<2.5$~kpc) halo stars (circles), and the 10\% fastest (diamonds); 
	bins of $5~\rm{km~s}^{-1}$ width are used. 
	The error bars are derived from Poisson's statistics of the counts. 
	{\bf Top middle-right:} 
	3D velocity distribution (detection space). 
	Filled dots show the $163$ sources with pairwise velocity differences less than $15~\rm{km\, s^{-1}}$. 
	Different colors indicate stars associated with the eight clumps recovered by the clustering analysis. 
        {\bf Bottom:} Toomre diagram of the full selected 
        sample, as above. 
	The 10\% fastest are marked with red-open dots. 
   	Members of GSE (x) and HS (+) are highlighted. 
  	The approximate location of known substructures 
	(GSE, HS, Nyx, Sequoia, the ``arc'', and Thamnos) 
	is shown. 
  	The (conservative) kinematical selection threshold for halo stars,  $\vert\vert{\bf v}-{\bf v}_{\rm LSR}\vert\vert> 230$ km s$^{-1}$, 
	is represented by the dashed line.   
          }
	\label{fig:fig2}
\end{figure*}


\section{Results}\label{sec:3}

Focusing on the sample of $1137$ stars to $2.5~\rm{kpc}$ of the Sun, 
we attempt detection and characterization of kinematic halo substructures in the solar vicinity as stars moving with similar space-velocities.

\subsection{Stream Detection}\label{sec:3.1}

The velocity distribution of the above sample is consistent with the superposition of stars belonging to the MW halo and metal-weak thick disk.

Following the two-point velocity correlation function 
methodology in \citet[][]{ReFiorentin2015}, 
we quantify deviations from the smooth kinematic distribution expected for the background population, possibly due to moving groups;
$K$-medoids cluster analysis in  
velocity space 
is then applied for final confirmation of substructures.

Figure~\ref{fig:fig2} (top left panel) shows 
$\xi({\bf \vert\vert{\bf v}_{\rm i}-{\bf v}_{\rm j}\vert\vert})$ 
for the full sample of $1137$ stars (circles), and
separately for the subsample of the 10\% fastest (diamonds). 
Clumping due to kinematical substructures (i.e., groups of particles moving coherently) is evident for values of $\langle \xi \rangle > 1$. 
There is a statistically significant signal 
for the full sample that peaks at $10-15~\rm{km~s}^{-1}$; 
it appears even stronger for the fastest subset. 

Next, we concentrate on the $163$ stars with velocity differences less than $15~\rm{km~s}^{-1}$; 
this number excludes stars generating isolated velocity-pairs to make the analysis more robust. 

We assign membership to these objects 
by using the implementation of the $K$-medoids clustering in 3D $(U,V,W)$ velocity space \citep[see e.g., ][]{Hastie2001} developed as part of 
the {\it R Project for Statistical Computing}: \verb+www.r-project.org+.

In order to get final identification, 
we compare 
runs with different $K$-classes and we choose the solution with the best Jaccard's index, $J$, that is an indicator of the stability of the groups \citep[][]{Tan}.

Gauging the similarity of the $K$-classes obtained for $100$ randomly selected half-samples, we find that $K=8$ maximizes $J$ for all the detected clusters (Table~\ref{table:1}).

In Figure~\ref{fig:fig2} 
we show, in detection space, these eight 
kinematical streams that, among the full sample, 
are visualized as filled dots with different colors (top middle-right);  
we also present the Toomre diagram (bottom panel). Here, 
the $10$\% fastest objects are highlighted with red-open dots, 
stars classified as GSE members by \citet[][]{Helmi2018} are marked with a times-sign, 
and Helmi Stream (HS) members, as found in \citet[][]{Koppelman2019HS}, with a plus-sign. 
We also 
show the approximate location\footnote{The regions shown are based on Fig. 2 in \citet[][]{Koppelman2019} as well as on known members of HS, GSE, Nyx listed in 
\citet[][]{Koppelman2019HS}, \citet[][]{Helmi2018}, \citet[][]{Necib2020}, respectively.} 
of known substructures (e.g., GSE, HS, Nyx, Sequoia, and Thamnos) 
which encompass the bulk of the high velocity stars and new GSE members from our  two retrograde groups.
 
Instead, the six new prograde streams clearly belong to the region usually associated with the Galactic disk, $\vert\vert{\bf v}-{\bf v}_{\rm LSR}\vert\vert< 230~\rm{km~s}^{-1}$.

\subsection{Chemo-dynamical properties}\label{sec:3.2}

\begin{deluxetable*}{lrrrrrrrrrrrrrrrrcccc}
\tablenum{1}
\tablecaption{\label{table:1} Chemo-dynamical mean Characteristics of the Eight Kinematical Groups.}
\tablewidth{0pt}
\tablehead{
\colhead{Group}  & N. &{\bf $J$}& \multicolumn{2}{c}{$\langle\rm{[Fe/H]}\rangle$}  & \multicolumn{2}{c}{$\langle\rm{[Mg/Fe]}\rangle$}  & \multicolumn{2}{c}{$\langle U\rangle$}& \multicolumn{2}{c}{$\langle V+V_{\rm{LSR}}\rangle$}& \multicolumn{2}{c}{$\langle W\rangle$}
& \multicolumn{2}{c}{$\langle L_{z}\rangle$} & \multicolumn{2}{c}{$\langle L_{xy}\rangle$} & \multicolumn{2}{c}{$\langle Z_{\rm max}\rangle$}  &
 \multicolumn{2}{c}{$\langle e\rangle$}\\
& & & \multicolumn{2}{c}{$\rm{(dex)}$} & \multicolumn{2}{c}{$\rm{(dex)}$} & \multicolumn{2}{r}{$\rm{(km~s^{-1})}$} & \multicolumn{2}{r}{$\rm{(km~s^{-1})}$} & \multicolumn{2}{r}{$\rm{(km~s^{-1})}$} & \multicolumn{2}{c}{$\rm{(kpc~km~s^{-1})}$} & \multicolumn{2}{c}{$\rm{(kpc~km~s^{-1})}$} & \multicolumn{2}{c}{$\rm{(kpc)}$} &  & \\
& & & \multicolumn{1}{c}{$\mu$}& \multicolumn{1}{c}{$\sigma$}&  \multicolumn{1}{c}{$\mu$}& \multicolumn{1}{c}{$\sigma$}& \multicolumn{1}{c}{$\mu$}& \multicolumn{1}{c}{$\sigma$}&  \multicolumn{1}{c}{$\mu$}& \multicolumn{1}{c}{$\sigma$}& \multicolumn{1}{c}{$\mu$}& \multicolumn{1}{c}{$\sigma$}& \multicolumn{1}{c}{$\mu$}& \multicolumn{1}{c}{$\sigma$}&  \multicolumn{1}{c}{$\mu$}& \multicolumn{1}{c}{$\sigma$}& \multicolumn{1}{c}{$\mu$}& \multicolumn{1}{c}{$\sigma$}&  \multicolumn{1}{c}{$\mu$}& \multicolumn{1}{c}{$\sigma$}
}
\startdata
$1$-orange & $10$ &$0.86$& $-1.36$ & $0.22$ &  $0.22$ & $0.06$ &$168$ &$31$ &$5$&$44$&$32$&$23$&       $-7$ & $379$ & $268$ & $132$ & $1.77$ & $0.93$ &  $0.87$ & $0.09$\\
$2$-cyan    & $17$ &$0.79$& $-1.50$ & $0.28$ &  $0.20$ & $0.16$ &$-82$&$49$&$88$&$37$&$40$& $16$& $783$ & $333$ & $377$ & $166$ & $1.78$ & $0.87$ &  $0.68$ & $0.14$ \\
$3$-pink     & $29$ &$0.74$& $-1.37$ & $0.28$ &  $0.14$ & $0.18$ &$-3$&$16$&$187$&$16$&$-7$& $25$& $1474$ & $148$ & $219$ & $100$ & $0.66$ & $0.47$ &  $0.23$ & $0.07$ \\
$4$-yellow$^{\star}$  & $44$ &$0.85$& $-1.45$ & $0.37$ & $-0.02$ & $0.25$ &$18$&$22$&$231$&$16$&$1$& $21$& $1875$ & 166$$ & $153$ & $140$ & $0.48$ & $0.59$ & $0.11$ & $0.05$ \\
$5$-blue     & $16$ &$0.81$& $-1.50$ & $0.37$ &  $0.20$ & $0.14$ &$64$&$25$&$131$&$33$&$53$& $28$& $1052$ & $288$ & $457$ & $230$ & $1.81$ & $1.23$ & $0.48$ & $0.16$ \\
$6$-green   & $18$ &$0.82$& $-1.43$ & $0.33$ &  $0.19$ & $0.15$ &$-62$&$18$&$176$&$18$&$-13$& $30$& $1425$ & $142$ & $268$ & $144$ & $1.01$ & $0.63$ & $0.33$ & $0.06$ \\
$7$-magenta & $19$ &$0.94$& $-1.56$ & $0.38$ &  $0.21$ & $0.19$ &$61$&$19$&$144$&$20$&$-45$& $19$& $1174$ & $150$ & $375$ & $167$ & $1.49$ & $0.91$  & $0.44$ & $0.09$ \\
$8$-red   & $10$ &$0.83$& $-1.37$ & $0.19$ &  $0.21$ & $0.08$ &$-139$&$55$&$-37$&$10$&$-42$& $80$&    $-285$ & $92$ & $548$ & $398$ & $3.55$ & $2.52$  & $0.76$ & $0.18$ \\
\enddata
\tablecomments{$^{\star}$This group is named ``Icarus" in this article. 
}
\end{deluxetable*}

\begin{figure*}
	\centering
	\gridline{\fig{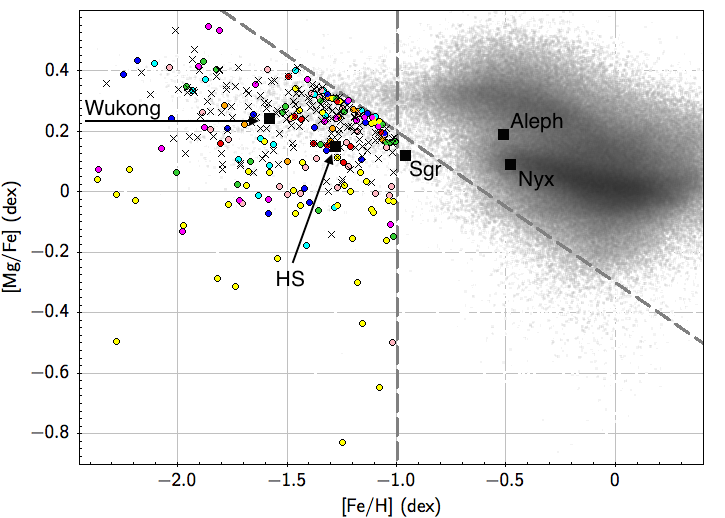}{0.5\textwidth}{}
                      \fig{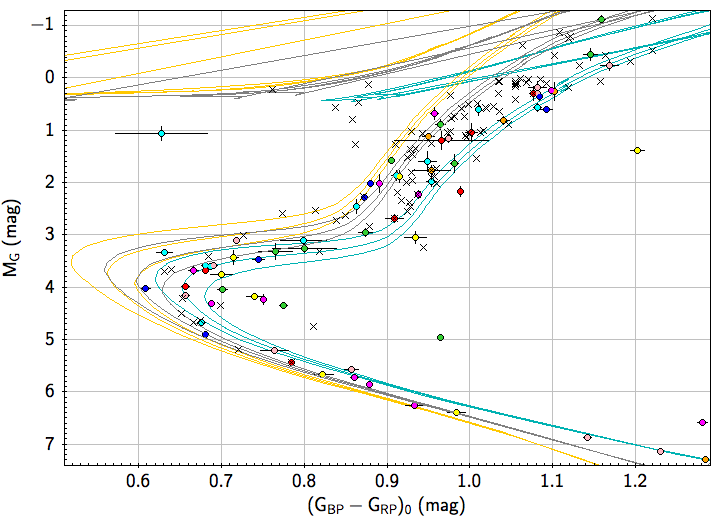}{0.5\textwidth}{}
                      }
	\caption{Distribution of the eight kinematical streams and GSE members as shown in Figure~\ref{fig:fig2}.	
     {\bf Left:} Chemical abundances, $\rm{[Mg/Fe]}$--$\rm{[Fe/H]}$. 
     In the background, all of the stars in our dataset. 
     Published values for the prograde structures Nyx, Aleph, Wukong, HS, and Sagittarius (Sgr) are highlighted with squares  
     \citep[][]{Necib2020, Naidu2020}. 
     {\bf Right:} CMD diagram ${\rm{M_G}}$ vs. ${\rm (G_{BP}-G_{RP})_0}$. 
     We only show the stars with $|b|>30^\circ$. 
     Isochrones of ages $10, 12, 14~\rm{Gyr}$ for $\rm{[M/H]}=-1.0$ (aqua), $\rm{[M/H]}=-1.5$ (silver), and $\rm{[M/H]}=-2.0$ (gold) 
     are from \citet[][]{MAWeiler2018}. 
          }
	\label{fig:fig3}
\end{figure*}

Table~\ref{table:1} lists mean values and dispersions of 
$\rm{[Fe/H]}$, $\rm{[Mg/Fe]}$,  $(U, V, W)$,  $(L_{z}, L_{xy})$, $Z_{\rm max}$, and eccentricity $e$. 
Also, Figure~\ref{fig:fig3} (left panel) shows the distribution $\rm{[Mg/Fe]}$--$\rm{[Fe/H]}$ for the full chemo-kinematical catalog, 
the members of the eight groups and the GSE objects. 
The {\it loci} of the prograde structures Nyx, Aleph, Wukong, HS, and Sagittarius are marked with their published values (squares). 

Group~1 (orange filled dots) and Group~8 (red) are slightly retrograde substructures, which appear associated with GSE: 
they have high eccentricity ($\langle e \rangle \gtrsim 0.7$) and $Z_{\rm max}$ (up to $7~\rm{kpc}$), confirming such debris are part of the accreted halo. 

Among our six prograde substructures, 
the most noticeable is Group~4 (yellow). It is  
a circular structure ($\langle e \rangle \lesssim 0.11$) 
confined to move close to the Galactic plane ($\langle Z_{\rm max} \rangle \lesssim 0.5$~kpc) with typical thin disk kinematics 
($\langle V+V_{\rm{LSR}}\rangle \simeq 231~\rm{km~s^{-1}}$ and $\sigma_V \simeq 16~\rm{km~s^{-1}}$). 
However, its chemical composition is not consistent with the abundances expected for the native MW thin disk. 
The low metallicity, $\langle\rm{[Fe/H]}\rangle \simeq -1.4$, and low $\alpha$-abundance, $\langle \rm{[Mg/Fe]}\rangle \simeq 0$ of Group~4 
indicate that these stars are most likely debris from an accreted satellite. 
To the best of our knowledge, this 
is the first detection of a flat prograde fast-rotating stream in the MW disk. Because of its characteristics, we name Group~4 
``Icarus"\footnote{
In the Greek mythology, Icarus is the son of Daedalus. He ignored his father's advice not to fly too close to the Sun and fell into the sea. 
Similarly, our Icarus stream appears to be originated from a dwarf galaxy that traveled too close to the MW;  
its debris are now fully spread in the ``ocean" of disk stars and seen to be flat and fast-rotating with the Sun.
}.

Group~3 (pink) and Group~6 (green) 
are characterized by quite flat and circular orbits with $\langle Z_{\rm max}\rangle \lesssim 1$~kpc and $\langle e\rangle\lesssim 0.3$.
Despite a rotation-velocity similar to the thick disk, $\langle V+V_{\rm{LSR}}\rangle \sim 180~\rm{km~s^{-1}}$, 
these groups are significantly more metal-poor ($\langle\rm{[Fe/H]}\rangle\simeq -1.4$); 
furthermore, they show intermediate $\alpha$-abundances ($\langle \rm{[Mg/Fe]}\rangle \lesssim +0.2$) 
that are not consistent even with the $\alpha$-enhanced metal-weak tail of the thick disk \citep[][]{Naidu2020}.
Such chemo-dynamical properties confirm that these groups must be debris of past merging events. 

An accreted origin 
is also expected for Group~5 (blue) and Group~7 (magenta), 
which show chemical compositions similar to 
Group~3 and 6 ($\langle\rm{[Fe/H]}\rangle\simeq -1.5$ and $\langle \rm{[Mg/Fe]}\rangle\simeq +0.2$). 
We argue that they are debris from two passages of the same satellite, 
as their mean angular momentum and eccentricity  are quite similar, while the vertical velocity components 
are close in modulus but opposite directions: 
$\langle W\rangle = +53~\rm{km~s^{-1}}$ and $-45~\rm{km~s^{-1}}$ for Group~5 and Group~7, respectively.
 
Finally, Group~2 (cyan) is a halo stream with 
mean velocity $(\langle U\rangle, \langle V+V_{\rm{LSR}}\rangle, \langle W\rangle)\simeq (-82, 88,40)~\rm{km~s^{-1}}$ and 
high-eccentricity $\langle e \rangle \simeq 0.68$. 
Its chemical composition similar to 
Groups~5, 6, and 7 ($\langle\rm{[Fe/H]}\rangle\simeq -1.5$ and $\langle\rm{[Mg/Fe]}\rangle\simeq +0.20$) 
confirms that it belongs to the accreted component of the MW halo. 

\subsection{Hints on the age}\label{sec:3.3}

Figure~\ref{fig:fig3} (right panel) shows members of the eight kinematical streams and GSE in the color magnitude diagram (CMD) 
${\rm{M_G}}$ vs. ${\rm (G_{BP}-G_{RP})_0}$. 
All stars have been corrected for extinction using the maps of \citet[][]{Schlafly}. 
Only objects with ${\rm (G_{BP}-G_{RP})_0}<1.3$ and $|b|>30^\circ$ are shown. 
Error bars include parallax and photometric errors, as well as extinction uncertainties. 
For reference, PARSEC-COLIBRI \citep[][]{Bressan2012, Marigo2017} isochrones\footnote{Trasmission curves are from \citet[][]{MAWeiler2018}.} 
with stellar age $10$, $12$, and $14~\rm{Gyr}$ and metallicities 
$\rm{[M/H]}=-1.0$ (aqua), $\rm{[M/H]}=-1.5$ (silver), and $\rm{[M/H]}=-2.0$ (gold) are also shown. 

A closer look at Figure~\ref{fig:fig3} (right panel) reveals some intriguing features. 
The color of the RGB stars shows significant dispersion, corroborating the spectroscopic evidence of a metallicity spread.
The bulk of RGB stars seems to be well constrained by our isochrones with $\rm{[M/H]}=-1.0$ and $-1.5$. In terms of age, stars in both our streams and GSE appear to be described by an old population. 
Indeed, if a metallicity  $\rm{[M/H]}=-1.0$ is adopted, the age range is between $10$ and $13~\rm{Gyr}$, whereas it is closer to $13~\rm{Gyr}$ if one adopts metal poorer isochrones. 
We point out that a non negligible fraction of the stars is redder than our most metal rich isochrones 
(with the exception of the star with  ${\rm (G_{BP}-G_{RP})_0} \approx 0.63$ and  ${M_{\rm G}} \approx 1$, 
which shows evidence of variability). These less extreme outliers may be due to unresolved binaries, 
enhancement in $\alpha$-elements (the PARSEC-COLIBRI isochrones have solar-scaled composition), underestimated chemical composition errors, 
lower accuracy of extinction estimates.

   \begin{figure*}
   \centering
    \includegraphics[width=\linewidth]{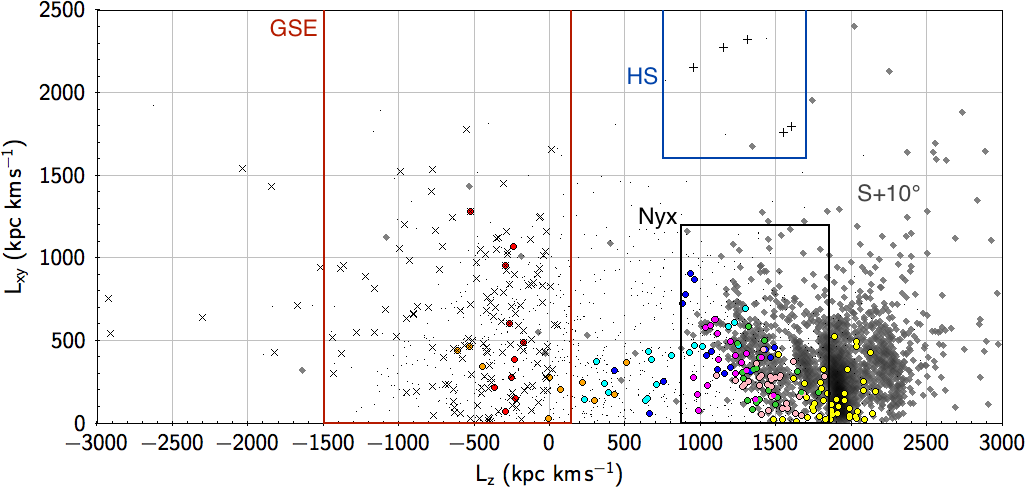}
   \caption{
	Space of adiabatic invariants, $L_{xy}$--$L_{z}$, for all the objects shown in Figure~\ref{fig:fig2},
	including the eight streams, GSE and HS members. 
	Solid boxes show the location of GSE, HS, and Nyx.  
	The debris 
	of the simulated $10^\circ$-inclination prograde satellite 
	analyzed in \citet[][]{ReFiorentin2015} are overplotted for comparison (diamonds).
	}
   \label{fig:fig4}
 \end{figure*}

\subsection{Comparison to simulations}\label{sec:3.4}

The analysis above has revealed two substructures associated with GSE 
and six new prograde streams. Their stars are old 
and their chemo-dynamical properties are clearly evidence of ancient mergers.

In order to better characterize the six co-rotating substructures, 
we compare our results to high-resolution N-body numerical simulations of minor\footnote{ 
$M_{\rm primary}/M_{\rm satellite} \sim 40$, similar to the estimated mass ratio of the MW relative to the Large Magellanic Cloud (LMC).} 
mergers published by \citet[][]{Murante2010} and 
used to study galaxy interactions and properties of accreted debris 
around the Sun by \citet[][]{ReFiorentin2015}. 

We explore the space of ``integrals of motion'' defined by the components of angular momentum in and out the Galaxy's disk. 
Figure~\ref{fig:fig4} shows the plane $L_{xy}$--$L_{z}$ 
for the full sample of $1137$ metal-poor stars chemically selected by means of Eq.~\ref{eq} (dots); 
stars associated with the eight lumps recovered by our cluster analysis in velocity space,  
as well as GSE and HS members, are visualized as before.  
Solid boxes show the {\it loci} of GSE, HS, and  Nyx. 
The $3585$ simulated stars from the $10^\circ$-inclination prograde satellite ($S+10^\circ$), 
 selected within a sphere of $2.5~\rm{kpc}$ centered at the ``Sun", 
 are overplotted (diamonds).
	
The consistency between the simulation and the prograde substructures is remarkable. 
In particular, Figure~\ref{fig:fig4} indicates that Icarus represents the debris of a low-inclination prograde satellite with a stellar mass 
$\sim 10^9  M_\sun$, similar to the LMC.
In fact, a massive satellite on a $10^\circ$-inclination prograde orbit, because of the efficient action of dynamical friction, 
quickly loses its orbital energy and circularize. 
Thus, it proceeds to the inner regions of the main halo, 
and leaves debris, stripped during multiple passages, with disk-like kinematics \citep[cfr. Figs. 6--7 by][yellow dots]{ReFiorentin2015}.

Groups~3, 5, 6, and 7 
might be either streams produced by previous orbital passages of the same progenitor of Icarus, 
or remnants from different 
satellites accreted along an initial prograde orbit with inclination $>+10^\circ$.
It is also plausible that these groups belong to Nyx and include its low-energy members; 
such accreted objects are more difficult to be separated from the {\it in situ} stars and could not be efficiently 
detected by the classification algorithm of \citet[][]{Necib2020}.

Finally, the nature of Group~2 remains uncertain, as it does not appear to be associated with either Icarus, HS or Nyx;  
it might represent the prograde tail of GSE. However, its progenitor should be a satellite on an intermediate-inclination, prograde or slightly retrograde, orbit.


\section{Discussion and Conclusions}\label{sec:4}

We have assembled a chemo-kinematical catalog based on top quality astrometric and spectroscopic data from Gaia~DR2, APOGEE~DR16, and GALAH~DR2. 
This data set can be exploited to explore a spherical volume around the  Sun up to $10~\rm{kpc}$.

We have chemically selected a sample of $1137$ 
stars to $2.5~\rm{kpc}$, and carried out statistical analysis and classification of their kinematics. 
Members of known substructures 
(e.g., HS, GSE, Sequoia, and Thamnos, by \citet[][]{Helmi1999, Belokurov2018, Helmi2018, Myeong2019, Koppelman2019}) are present in the 10\% fastest subsample.

Among the subsample of $163$ objects with relative velocity less than $15~\rm{km~s}^{-1}$, 
we have found statistical evidence of eight kinematical substructures.
The low $\alpha$-abundances 
of their members is quite consistent with the low-metallicity tail of a progenitor dwarf galaxy 
similar to the LMC \citep[][]{Nidever2020}.
Also, comparing their CMD to PARSEC-COLIBRI isochrones, 
these substructures appear to be older than $10~\rm{Gyr}$. 

The two retrograde groups are associated with GSE; 
while the six prograde substructures are located in a region that was difficult to explore with the halo sample selection criteria traditionally applied. 
We have further investigated their origin by means of comparison to high-resolution N-body numerical simulations 
of the interaction between a MW-like galaxy and orbiting LMC-like dwarf galaxies \citep[][]{Murante2010}. 

Most noticeable, the new Group~4, that we named Icarus, 
is the ``flattest" among the fast-rotating streams previously found in the Galactic disk. 
Clearly, the stellar ages greater than $10~\rm{Gyr}$ rule out the possibility that this kinematical group is formed by in situ disk stars. Instead, 
its peculiar chemo-dynamical properties 
($\langle \rm{[Fe/H]}\rangle \simeq -1.45$, $\langle \rm{[Mg/Fe]}\rangle \simeq -0.02$, and $\langle e \rangle \simeq 0.11$) 
are consistent with debris from a dwarf galaxy progenitor with a stellar mass of 
$\sim 10^9  M_\sun$ on an initial prograde very low-inclination orbit. 
We notice that it shares dynamical properties similar to Aleph, the metal-rich stream ($\langle \rm{[Fe/H]}\rangle \simeq -0.5$), 
discovered outside the plane ($|Z|>2$~kpc) 
by \citet[][]{Naidu2020} and extending up to $10~\rm{kpc}$.

It is plausible that Groups~3, 5, 6, 7 are either streams previously released by the same progenitor of Icarus, 
or remnants from different 
satellites accreted along an initial prograde orbit, but with inclinations $>+10^\circ$.
These debris could also be low-energy members of Nyx \citep[][]{Necib2020}.

As for Group~2, the high eccentricity and low angular momentum ($\langle e \rangle \simeq 0.7$, $\langle L_{xy}\rangle \simeq 377~\rm{kpc~km~s^{-1}}$) 
exclude its association with any of Icarus, HS and Nyx.
It is chemically similar to Wukong \citep[][]{Naidu2020} and GSE; if they could have a common origin, 
Group~2 would represent debris from a more recent passage of the Wukong progenitor or the prograde tail of GSE.

Future work will have to disentangle on the common origin of these streams, based on even better data from the next releases of Gaia and 
in continuous synergy with ground-based spectroscopic surveys. 


\acknowledgments
We are grateful to the Anonymus Referee for comments that helped us improve the original manuscript. 
We wish to thank A. Curir, E. Poggio, and R.~L. Smart for helpful discussions. 
M.~C. acknowledges the support of INFN ``Iniziativa specifica TAsP". 
We are indebted to the Italian Space Agency (ASI) for their continuing support through contract 2018-24-HH.0 to the National Institute for Astrophysics (INAF).
This work has made use of data from the European Space Agency (ESA) mission Gaia (https://www.cosmos.esa.int/gaia), 
processed by the Gaia Data Processing and Analysis Consortium (DPAC, https://www.cosmos.esa.int/web/gaia/dpac/consortium). Funding for the DPAC has been provided by national institutions, in particular the institutions participating in the Gaia Multilateral Agreement. 


{}

\end{document}